\input epsf.tex    

\documentstyle[12pt]{article}

\newcommand{\bmat}{\left(\begin{array}}
\newcommand{\emat}{\end{array}\right)}

\def\yzero{\smash{\hbox{$y\kern-4pt\raise1pt\hbox{${}^\circ$}$}}}

\def\beq{\begin{equation}}
\def\eeq{\end{equation}}
\def\beqa{\begin{eqnarray}}
\def\eeqa{\end{eqnarray}}

\def\-{\hphantom{-}}
\def\ov{\overline}
\def\s2{\frac{1}{2}}

\def\beq{\begin{equation}}
\def\eeq{\end{equation}}
\def\beqa{\begin{eqnarray}}
\def\eeqa{\end{eqnarray}}

\def\IF{\relax{\rm I\kern-.18em F}}
\def\II{\relax{\rm I\kern-.18em I}}

\def\cp{{\cal P}}
\def\IC{\bf C}
\def\IZ{\bf Z}
\def\IR{\bf R}
\def\IX{\bf X}
\def\IS{\bf S}
\def\IP{\bf P}
\def\z2z2{$\IC^3/(\IZ_2\times\IZ_2)$}

\def\Dsl{\,\raise.15ex\hbox{/}\mkern-13.5mu D} 

\def\IT{\bf T}

 \def\cp#1{\relax\ifmmode {\IP\kern-2pt{}_{#1}}\else $\IP\kern-2pt{}_{#1}$\=fi}

\begin{document}

\makeatletter \@addtoreset{equation}{section} \makeatother
\renewcommand{\theequation}{\thesection.\arabic{equation}}
\pagestyle{empty}
\vspace*{.5in}
\rightline{FT-UAM-02-09}
\rightline{IFT-UAM-CSIC-02-08}
\rightline{\tt hep-th/0204079}
\vspace{2cm}

\begin{center}
\LARGE{\bf Localized instabilities at conifolds\\[10mm]}
\medskip
\large{Angel M. Uranga \footnote{\tt Angel.Uranga@uam.es} \\[2mm]}
Dpto. de F\'{\i}sica Te\'orica C-XI and 
Instituto de F\'{\i}sica Te\'orica C-XVI \\
Universidad Aut\'onoma de Madrid, 28049 Madrid, Spain \\ [3mm]

\smallskip

\small{\bf Abstract} \\[3mm]
\end{center}

\begin{center}
\begin{minipage}[h]{14.5cm}
{\small 
We consider the M-theory lifts of configurations of type IIA D6-branes 
intersecting at angles. In supersymmetry preserving cases, the lifts 
correspond to special holonomy geometries, like conifolds and $G_2$ 
holonomy singularities. Transitions in which D6-branes approach and 
recombine lift to topology changing transition in these geometries. In 
some instances non-supersymmetric configurations can be reliably lifted, 
leading to the same topological manifolds, but endowed with 
non-supersymmetric metrics. In these cases the phase transitions are 
driven dynamically, due to instabilities localized at the 
singularities. Even though in non-compact setups the instabilities relax 
to infinity, in compact situations there exist nearby minima where the 
instabilities dissappear and the decay reaches a well-defined (in general 
supersymmetric) endpoint.
}

\end{minipage}
\end{center}
\newpage
\setcounter{page}{1} \pagestyle{plain}
\renewcommand{\thefootnote}{\arabic{footnote}}
\setcounter{footnote}{0}

\section{Introduction}
\label{intro}

Recently new insights into the dynamics of non-supersymmetric string
configurations have been achieved by studying localized instabilities.  
For instance, open string tachyons localized on D-brane world-volumes (see
e.g. \cite{sen}), and closed string tachyons at twisted sectors of
non-supersymmetric orbifolds \cite{aps,dabholkar,vafa,others} 
\footnote{See \cite{fluxbranes} for a similar kind of localized tachyons 
in fluxbrane backgrounds. For some results on not localized closed 
tachyons, see e.g. \cite{fullclosed}.}

In this paper we study a new (but somewhat related) kind of instabilities, 
localized at certain non-supersymmetric singularities, obtained as U-duals 
of brane configurations (basically, as M-theory lifts of type IIA 
intersecting D6-brane configurations; a related kind of lifts has been 
recently studied in \cite{rauljoan}). The instabilities arise in the 
singular world-sheet CFT regime, a fact which makes their analysis 
difficult. In particular, standard tools like use of D-brane probes 
\cite{aps}, mirror symmetry \cite{vafa}, etc. are not valid. Instead, we 
rely on the intuition gained from the U-dual D-brane system, plus 
analysis of the energetics justified by use of BPS formulae.

Our singularities contain a set of continuous parameters (not dynamical 
moduli) connecting them to special holonomy metrics, for instance 
conifold singularities. The susy cases contain moduli which parametrize 
smoothing of the singularities, and/or topology changing transitions. Our 
main result is that in the non-supersymmetric situation these processes 
are dynamically triggered by the localized instability. In non-compact 
situations, the dynamical smoothing of the singularity is reminiscent of 
\cite{aps}. In compact setups
\footnote{Configurations of intersecting D6-branes in compact manifolds 
have been studied in \cite{orbif,intersecting,susy}.}, however, our 
localized instabilities present new features, namely they can reach a
stable point at finite distance, and they lead to no reduction in 
spacetime dimension. 

The paper is organized as follows: In Section \ref{twoangle} we describe 
configurations of D6-branes intersecting over five dimensions, and their 
M-theory lift given by conifold singularities. In Sections \ref{instab}, 
\ref{local} we argue that the configurations suffer an instability, and 
propose a dynamical conifold transition as the natural relaxation process. 
Section \ref{onethree} contains similar discussions for other D6-brane 
intersections, involving a dynamical topology changing flop of 2-spheres
(Section \ref{oneangle}) and a dynamical resolution of codimension seven
singularities, of $G_2$ holonomy in the supersymmetric case (Section
\ref{threeangle}). Section \ref{conclusions} contains our final 
comments. Appendix \ref{slag} reviews the construction of recombined 
special lagrangian cycles \cite{angle}, and appendix \ref{kahler} 
discusses topological obstructions to phase transitions from the D6-brane 
and M-theory viewpoints.

\section{Two angle system}
\label{twoangle}

\subsection{System of D6-branes intersecting over 5d}
\label{twodsix}

In this section we consider the dynamics of two D6-branes intersecting 
over a five-dimensional subspace of their world-volume. Without loss of 
generality, the geometry of the configuration is 
\beqa
{\rm D6}\;\; 0\; 1\; 2\; 3\; 4\; \times \; [\; 6\; 7\; ]_{\theta_1}\; [\; 
8\; 9\; ]_{\theta_2} \nonumber \\
{\rm D6}'\;\, 0\; 1\; 2\; 3\; 4\; \times [\; 6\; 7\; ]_{\theta_1'}\; [\; 
8\; 9\; ]_{\theta_2'} 
\eeqa
meaning that the D6-brane spans the directions 01234 times a line at angle 
$\theta_1$ with the $x^6$ axis in the $\IR^2$ parametrized by 67, times a 
line at angle $\theta_2$ with the $x^8$ axis in $({\bf R^2})_{89}$. 
Analogously for the D6$'$-brane.

We will also consider compact examples with 6789 parametrizing a $\IT^4$, 
taken rectangular for simplicity $(\IT^2)_{67}\times (\IT^2)_{89}$. The 
angles are then defined in terms of the torus radii $R_i$ and the wrapping  
numbers $(n_6,n_7,n_8,n_9)$, which moreover specify the homology class of 
the D6-brane worldvolumes
\beqa
[\, \Pi_{D6}\, ] \, = (\, n_6\, [a_1] \, + \, n_7 \, [b_1]\, ) \, \otimes 
\, (\, n_8\, [a_2]\, +\, n_9\, [b_2]\,)
\label{homol2}
\eeqa

The system has two branches: i) Since the direction 5 is transverse to 
both branes, they may be separated in that direction, their distance being 
controlled by the vev of a tree-level modulus $\rho$ (with radiative 
potential generated in non-supersymmetric situations, see below) neutral 
under the D6-brane gauge symmetries. We call this the Coulomb branch; ii) 
At the origin of the Coulomb branch, i.e. for intersecting branes, charged 
massless or tachyonic scalars $\phi$ arise at their intersection. Vevs 
for these fields parametrize a Higgs branch where both branes recombine 
into a single smooth one.

Let us define $\Delta \theta_i=\theta_i-\theta_i'$. The configuration is 
supersymmetric when the $SO(4)$ rotation relating the two 2-planes in 
$({\bf R^4})_{6789}$ is within $SU(2)$ \cite{bdl}, that is
\beqa
\Delta\theta_1\, \pm \, \Delta\theta_2\, =\, 0 \; ,\; {\rm for} 
\; {\rm some}\; {\rm choice}\; {\rm of}\; {\rm sign}, 
\eeqa
and if so preserves $1/4$ of the supersymmetries. In such situation the 
above  two branches are degenerate and the fields $\rho$ and $\phi$ are 
exact moduli (even non-perturbatively, due to the eight unbroken susys). 
For any non-zero vev for $\phi$, the system corresponds to a recombined 
D6-brane wrapped on a supersymmetric 2-cycle in the coordinates 6789. In 
the compact case its homology class is the sum of the original homology 
classes (\ref{homol2}) $[\Pi_{\rm tot}] = [\Pi_{\rm D6}] + 
[\Pi_{{\rm D6}'}]$; in the non-compact case, its asymptotic form is 
that of the original intersecting 2-planes in $\IR^4$. This 2-cycle is 
special lagrangian in the complex structure natural in our splitting 
$\IT^2\times \IT^2$. In a different complex structure it is a holomorphic 
2-cycle in the right homology class. 

In other cases, $\Delta\theta_1\, \pm \, \Delta\theta_2\, \neq \, 0$ for 
both signs, all supersymmetries are broken, and moduli space is lifted. 
For instance, for $0\leq \Delta \theta_i \leq \pi$, the lightest scalar 
fields arising at the intersection of branes at the origin of the Coulomb 
branch have masses
\beqa
\alpha' \, M^2 = \frac 1{2\pi} (\Delta\theta_1 - \Delta\theta_2) \quad 
\quad ; 
\quad \quad
\alpha' \, M^2 = \frac 1{2\pi} (-\Delta\theta_1 + \Delta\theta_2) 
\eeqa
In any non-supersymmetric situation, $\Delta \theta_1 - \Delta\theta_2
\neq 0$, exactly one of these complex scalars is tachyonic. In the 
supersymmetric case, both get massless and combine with fermions to fill 
out a hypermultiplet. 

The tachyon at the intersection triggers the dynamical recombination of 
the D6-branes. Since the tachyon is localized, the decay proceeds via an 
expanding shell which leaves the recombined configuration behind.
In the non-compact case, a simple picture of the shell at any finite time 
is as follows. Outside a 3-ball of finite extent the D6-branes are 
unperturbed and span two 2-planes (minus two disks whose interior is 
inside the shell) at angles; inside the 3-ball the recombined cycle is a 
special lagrangian 2-cycle with boundary given by two disks glued to the
outside solution at the shell. The existence and explicit construction of 
this configuration is provided in \cite{angle}, see appendix \ref{slag}. 
The difference in tension dynamically pushes the joining shell to infinity. 
The process can be regarded as triggered by the intersecting tachyon for 
small $\phi$ vevs, and by the Dirac-Born-Infeld action (which tends to 
minimize the volume of the recombined cycle) deep in the Higgs branch.

In the non-compact case the process proceeds to infinity. In the compact 
case, the D6-brane ends up wrapping the minimal volume supersymmetric 
2-cycle in the class $[\Pi_{\rm tot}]$. This configuration can be regarded 
as the minimum of the tachyon potential (possibly multi-tachyon potential, 
if there are multiple intersections \footnote{In compact examples, there 
may be obstructions to such recombinations, for instance in cases with a 
single intersection point, see appendix \ref{kahler}. In the remainder of 
the paper we consider the relevant transitions to be allowed.}).

Concerning the Coulomb branch, two D6-branes at angles suffer a mutual 
attraction generated by tree-level exchange of closed string fields, or 
equivalently by a one-loop running of open string states. In any event, a 
potential is developed for $\rho$ which pushes the system towards the 
origin of the Coulomb branch. Hence, two D6-branes initially deep in the 
Coulomb branch dynamically tend to approach and suffer a transition to a 
Higgs branch, where they recombine. This process has been described in 
\cite{raulcosmo} (in a particular application as a model for hybrid 
inflation; see \cite{herdeiro} for an earlier proposal to use Coulomb to 
Higgs transitions in brane models as hybrid inflation scenarios).

\subsection{Energetics in the M-theory lift}
\label{instab}

Since the M-theory lifts of systems of D6-branes are purely geometrical, 
we would expect the above system to provide interesting dynamical 
phenomena involving purely gravitational systems. Our purpose in this 
section is to support the existence of the above Coulomb to Higgs 
transition in the IIA strong coupling regime (M-theory supergravity regime), 
and analyze its geometrical interpretation in M-theory.

The M-theory lift of $n$ D6-branes is given by $\IR^7\times \IX_4$ 
where $\IX_4$ is an $n$-center Taub-NUT geometry, with metric given by
\beqa
& ds_{TN}^2 \, = \, V\, d{\vec r}^{\,2}\, +\, V^{-1}\, (\, d\psi\, +\, 
{\vec \omega}\, \cdot\, d\vec{r}\,)^2 & \nonumber\\
& V\, =\, \frac 1R\, +\, \sum_{i=1}^n\, \frac{1}{|{\vec r} - {\vec{r_i}}|}\,
\quad \quad ; \quad \quad  \vec{\nabla}\, \times\, \vec{\omega}\, =\, 
\vec{\nabla}\, V &
\eeqa
The metric describes an $\IS^1$ fibration over $\IR^3$, parametrized by 
$\vec{r}$, with fiber degenerating over the locations $\vec{r_i}$, and 
having asymptotic constant radius $R$. In the limit of large asymptotic 
radius (equivalently, in the near core region), the constant term in $V$ 
drops, and the metric becomes asymptotically conical, an ALE geometry.

The naive M-theory lift of two D6-branes intersecting as in section 
\ref{twodsix} corresponds to a six-dimensional geometry looking like two 
intersecting Taub-NUT fibrations. The full-fledged metric of these systems 
have not been studied, but their topology is relatively clear. As we will 
discuss below (see Section \ref{local}) in the supersymmetric situation 
the constraints from supersymmetry are enough to fix the metric in the 
infinite asymptotic radius regime (or near core limit) to be the conifold 
metric. Hence, we expect the dynamics of the system in non-supersymmetric 
situations to teach us about dynamics of non-supersymmetric metric 
deformations of the conifold.

\medskip

The first question that arises is how one may extrapolate to strong 
coupling the picture we obtained studying D6-brane systems at weak 
coupling. A realiable way to do so is to use BPS formulae for the tension 
of the system, in situations where supersymmetry is good enough to prevent 
strong corrections from appearing.

Let us consider the compactification of M-theory on a rectangular 5-torus, 
$\IT^5=\IT^2\times \IT^2\times \IS^1$, parametrized by 67, 89, 10 
respectively. Consider a state given by the superposition of 
two far-away Kaluza-Klein monopoles associated to the $\IS^1$ direction 
$10$. One of them spans the directions 01234 and wraps the cycles 
$(n_6,n_7)$ and $(n_8,n_9)$ in $\IT^2\times \IT^2$; the second spans 01234 
and wraps the cycles $(n_6',n_7')$ and $(n_8',n_9')$. Both are separated 
by a large distance along the direction 5. Due to factorization, for 
distances much larger than the M-theory circle the interactions between 
the two objects are negligible and the state is reliably represented by a 
superposition of two Taub-NUT geometries. In non-compact space the metric 
would be roughly speaking of the kind
\beqa
&ds \, = \, d\vec{x}_{01234}^{\,2}\, +\, (V+V')\, d{\vec r}^{\,2} \, +\, 
(V+V')^{-1}\, (\, d\psi\, +\, {\vec \omega}\, \cdot\, d\Delta\vec{r}\,
+\, {\vec \omega'}\, \cdot\, d\Delta\vec{r}'\, )^2 & \nonumber\\
& V\, =\, \frac 1{2R}\, +\, \frac{1}{|\Delta {\vec r}|} \quad \quad ; 
\quad \quad
\vec{\nabla}\, \times_{3}\, \vec{\omega}\, =\, \vec{\nabla}\, V &
\eeqa 
where 
\beqa
\Delta {\vec r}\, =\, (\, x_5-x_0\, ,\, -\sin \theta_1 x_6 + \cos \theta_1 
x_7\, ,\, -\sin \theta_2 x_8 + \cos \theta_2 x_9\,) 
\eeqa
is the distance of a point to the TN center, and where $\times_{3}$ is 
vector product acting in the 3-dimensional space transverse to the TN 
center. Analogously for quantities $V'$, $\omega'$, $\times_3'$ 
associated to the second Kaluza-Klein monopole. 

The above metric does not solve the equations of motion, but is 
approximately correct for large separation $x_0-x_0' \gg$. This is as 
expected for weakly interacting KK monopoles.

\medskip

For well separated objects, the total tension $Z_{\rm tot}$ of the 
resulting five-dimensional object is just the sum of the individual KK 
monopole tensions. For future convenience, we define $q_i=R_i n_i$ and 
obtain
\beqa
Z_{\rm total} \, =\, Z_{q_i}\, +\, Z_{q_i'}
\eeqa
where
\beqa
Z_{q_i}\, =\, T_{KK}\, (\, q_6^2\, +\, q_7^2\, )^{1/2} 
\,\, (\,q_8^2\, +\, q_9^2\, )^{1/2}
\eeqa
and analogously for primed quantities. Here $T_{KK}= M_P^9\, R_{10}^{\,2}$ 
is the KK monopole tension. 

The state of KK monopoles at angles is generically non-supersymmetric, and 
does not saturate the BPS bound. Gravitational and electromagnetic 
interactions between two separated KK monopoles with differently oriented 
world-volume generically do not cancel and lead to an attractive interaction, 
pulling them towards the origin of the Coulomb branch.

Let us compute the BPS bound for a state with the above charges. Define 
the independent charges $q_{ij}=q_iq_j$, ${\tilde q}_{ij}=q_{ij}+q_{ij}'$, 
associated to a basis of homology cycles in $\IT^4$. The BPS bound can be 
analyzed as in e.g. \cite{op} and follows from the maximal 
eigenvalue of the central charge matrix, which for states with only 
(arbitrary) KK monopole charges reads. 
\beqa
Z = T_{KK} R_m R_n q_{mn} \Gamma^{mn} 
\label{bound}
\eeqa
where $\Gamma^{m}$ are Dirac matrices. In our case
\beqa
Z\, =\, T_{KK}\, (\, q_{68}\, \Gamma^{68}\, + \, q_{69}\, \Gamma^{69} \,+ 
\, q_{78} \, \Gamma^{78} \, +\, q_{79} \, \Gamma^{97}\, )
\label{bound2}
\eeqa

Since the total charges do not satisfy the quadratic constraint 
${\tilde q}_{68} \, {\tilde q}_{79}={\tilde q}_{69}\, {\tilde q}_{78}$ 
(except for the trivial case of collinear charges, namely parallel branes)
the BPS saturating state is at most 1/4 supersymmetric. The bound (for  
some assumed signs for the ${\tilde q}_{mn}$) is
\beqa
Z\, =\, T_{KK}\, [\, (\,{\tilde q}_{78} \pm {\tilde q}_{69}\, )^2\, +\, 
(\,{\tilde q}_{79} \mp {\tilde q}_{68}\,)^2\, ]^{1/2}
\eeqa
In terms of the angles $\theta_i$, $\theta_i'$
\beqa
q_6\, =\, (\, q_6^2\, +\, q_7^2\, )^{1/2}\, \cos\theta_1 \quad ; \quad 
q_8\, =\, (\, q_8^2\, +\, q_9^2\, )^{1/2}\, \cos\theta_2 \\
q_7\, =\, (\, q_6^2\, +\, q_7^2\, )^{1/2}\, \sin\theta_1 \quad ; \quad 
q_9\, =\, (\, q_8^2\, +\, q_9^2\, )^{1/2}\, \sin\theta_2
\eeqa
(and analogously for primed charges) and after some algebra, the BPS bound 
can be recast as
\beqa
Z\, =\, [\, (\, Z_{q_i}\, +\, Z_{q_i'}\, )^2\, -\, 4\, Z_{q_i} \, 
Z_{q_i'}\, \sin^2(\delta/2)\, ]^{1/2}
\eeqa
where $\delta\,=\, \Delta\theta_1- \Delta\theta_2$.

The order parameter $\delta$ vanishes when the relative angles define an 
$SU(2)$ rotation, in which case the BPS bound is $ Z_{q_i} + Z_{q_i'}$ and 
is saturated by the original two KK monopole configuration, which is 
supersymmetric. For non-zero $\delta$, the BPS bound is smaller, and the 
initial state is non-supersymmetric. At small M-theory radius, the bound is 
saturated by a KK monopole wrapped in the holomorphic curve in the class 
$[\Pi_{\rm tot}]$ in $\IT^4$. Since this state is supersymmetric, it 
persists even at large M-theory radius. In the supergravity regime it 
should look like a Taub-NUT space fibered over the 2-cycle. An `adiabatic' 
anstaz for the metric, by fibering the TN metric over the 2-cycle, would be 
reliable for $R_i\gg R_{10}$ (curvature radii for the 2-cycle much larger
than the $\IS_1$ compactification length). 

Hence the original state of two well separated KK monopoles is unstable 
against decay to the 1/4 BPS state which corresponds to the lift of a 
D6-branes wrapped on the recombined 2-cycle. On the Coulomb branch the 
instability is driven by the long-distance interaction between well 
separated KK monpoles. At Planckian distances, a local instability must 
develop in the region near the intersection of the two Taub-NUT cores, and 
triggers the transition to the final recombined state. In the following 
section we describe the local process mediating this decay to a Higgs 
branch.

\subsection{The dynamical conifold transition}
\label{local}

One particular advantage of the system we are stuying is that it contains 
a set of continuous parameters connecting it to a supersymmetric 
situation. In particular, one may study the lift to M-theory of the 
transition from the Coulomb to the Higgs branch in the supersymmetric 
configuration, and then perturb it mildly by a small change in the torus 
radii violating the $SU(2)$ relation between angles. In this fashion, the 
main effect of lack of supersymmetry is that the transition is dynamically 
triggered, instead of taking place along flat directions in moduli space.

In the supersymmetric situation, the phase transition in which D6-branes 
at angles approach and recombine lifts to a topology changing transition 
between two distinct Calabi-Yau threefolds. More concretely, the 
intersections lift to conifold singularities, and the phase transtition 
maps to a conifold transition from the small resolution to the deformation 
phase. Let us describe this story.

Consider two D6-branes intersecting over 01234 and spanning two 2-planes 
in 6789 in a supersymmetric fashion. By choosing suitable complex 
coordinates $z$, $w$, the locus in $\IR^4$ wrapped by the D6-branes may be 
written $z\,w\,=\,0$. Now recall that the M-theory lift of a D6-brane 
may be written as a complex manifold as $xy=v$; namely, a fibration 
of $\IC^*$'s (parametrized by $x$, $y$) over $\IC$ (parametrized by $v$) 
with fiber degenerating over $v=0$ (the location of the Taub-NUT center). 
Extending to our case, the M-theory lift of two D6-branes intersecting in 
$SU(2)$ angles can be described (as a complex manifold) as a $\IC^*$ 
fibration over $\IC^2$ (parametrized by $z$, $w$) degenerating over $z\, 
w\, = \, 0$. The total space is the submanifold $x\, y - z\, w = 0$ in 
$\IC^4$. This is the description of the conifold singularity as a 
complex manifold\footnote{See \cite{radu1} for a similar derivation. A 
different but related derivation starts with intesecting NS-fivebranes (a 
system U-dual to intersecting D6-branes) and obtains a conifold geometry 
by applying T-duality \cite{uraconi,dm}.}. Finally, supersymmetry of the 
configuration guarantees the threefold is endowed with the Calabi-Yau metric.

\begin{figure}
\begin{center}
\centering
\epsfysize=3cm
\leavevmode
\epsfbox{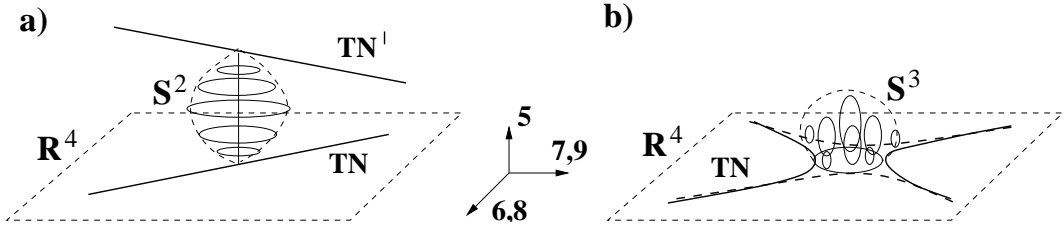}
\end{center}
\caption[]{\small 
Schematic picture of the lift of configurations of D6-branes intersecting 
at two angles on the Coulomb (a) and Higgs (b) branch. They are related by 
a topology changing conifold transition. We have highlighted the 
non-trivial two- and three-spheres in these geometries, which are obtained
as $\IS^1$ fibrations over a segment (a) and a disk (b) on the base.}
\label{conifold}
\end{figure}

The Coulomb branch in which the D6-branes separate corresponds to the 
small resolution phase of the conifold, where the singularity is replaced 
by a 2-sphere. In the lift of two D6-branes on the Coulomb branch, the 
2-sphere is clearly visible as the $\IS^1$ fibration over a segment  
joining the centers of the two Taub-NUTs on the base, see 
Fig~\ref{conifold}a.
On the Higgs branch, where the intersecting D6-branes recombine and wrap a 
single smooth 2-cycle, the system lifts to a $\IC^*$ fibration 
degenerating over the complex curve $z\,w\, = \, \varepsilon$, with 
$\varepsilon$ parametrizing the Higgs branch. The complex equation for 
the threefold is $ z\,w\, -\,x\, y\, = \, \varepsilon$, namely the 
deformed conifold, where the singularity is replaced by an $\IS^3$. In the 
lift of the recombined D6-brane, the $\IS^3$ is visible as the $\IS^1$ 
fibration over a disk on the base, bounded by a non-trivial circle in
the 2-cycle \footnote{If $\epsilon=r^2 e^{i\theta}$, the circle in 
$z_1 z_2=\epsilon$ is given by defining $z_1=e^{i\theta/2} (x_1+ i x_2)$, 
$z_2=e^{i\theta/2} (x_1- i x_2)$, and taking $x_i$ real in 
$x_1^2+x_2^2=r^2$.}, see Fig~\ref{conifold}b. 

Hence the transition from the Coulomb to the Higgs branch in which 
D6-branes approach and recombine lifts to the familiar conifold 
transition, in which a resolved conifold shrinks its two-sphere and 
instead a three-sphere grows \cite{candelas,gms}. The transition is a 
Higgs mechanism triggered by a vev for a state given by an M2-brane 
wrapped on the vanishing 2-cycle. This state is the lift of the 
hympermultiplet arising at the intersection of D6-branes.

\medskip

Performing a mild perturbation of the above system by a small change of 
angles away from the $SU(2)$ relation leads to presumably small 
corrections to the above picture. The topology of the resulting lift is 
unchanged, but the asymptotic behaviour of the lift (in the non-compact 
setup) forces the metric to be non-supersymmetric.

The main effect is that the conifold transition is driven by a dynamical 
mechanism which tends to minimize the energy and make the total tension of 
the system approach the BPS bound. Deep in the `Coulomb branch' the 
gravitational dynamics between the Taub-NUT cores leads to a dynamical 
shrinking of the 2-sphere. The nature of the instability arising near the 
origin in the Coulomb branch is less clear; a putative tachyonic nature of 
the wrapped M2-brane state, as naively extrapolated from the weak 
coupling IIA limit, may not be the right answer. 

\begin{figure}
\begin{center}
\centering
\epsfysize=3cm
\leavevmode
\epsfbox{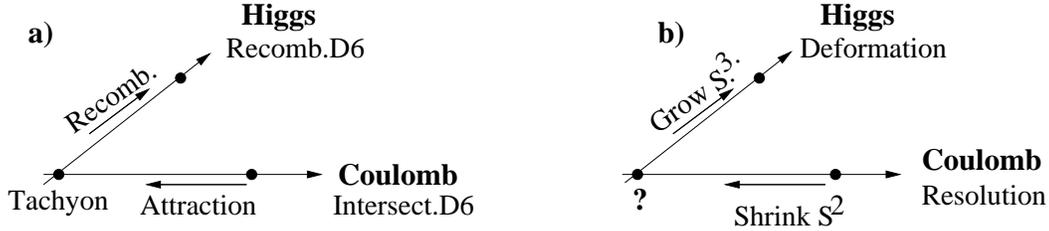}
\end{center}
\caption[]{\small Phase transition in the M-theory lift of intersecting 
D6-branes a) in the small circle limit (weakly coupled IIA) and b) in the 
large circle regime. Even though the nature of the instability is not 
fully understood (lies beyond supergravity due to small cycles) it 
certainly mediates a dynamical conifold transition transforming an 
initially large 2-cycle into a finally large 3-cycle.}
\label{diagram}
\end{figure}       

In any event, some M-theory mechanism triggers the transition to the Higgs 
branch. Further indirect evidence for this is the fact that, deep in the 
Higgs branch, for curvature lengths on the base cycle much larger than 
$R_{10}$, the dynamics is governed by the effective action for KK 
monopoles 
wrapped on the recombined cycle. Since it becomes mainly the  
Dirac-Born-Infeld action (see e.g. \cite{tomas, imamura}), it tends to 
minimize the wrapped volume, which corresponds to continuing the 
recombination. The complete situation is depicted in Figure \ref{diagram}.

We conclude this section giving two intuitive arguments to understand why 
the non-supersymmetric conifold geometry likes to deform dynamically. 
First one can resort to an analysis of the effective field theory for the 
$U(1)$ gauge field arising from integrating the 3-form over the vanishing 
2-cycle, and the charged `hypermultiplet' obtained from the M2-brane 
wrapped on it. In the supersymmetric situation, the conifold transition is 
a Coulomb to Higgs transition in this field theory (we understand additional 
conifolds to be present in compact setups, see appendix \ref{kahler}), via 
flat directions. The D6-brane angle rotation breaking supersymmetry breaks 
the $SU(2)$ R-symmetry of the system, hence it naturally corresponds to 
turning on a Fayet-Iliopoulos term in the conifold field theory. This 
naturally triggers a dynamical transition into the Higgs branch.

A second argument relies on partial results from T-duality between 
(unfortunately smeared) NS-brane configurations and conifolds. The 
conifold metric \cite{candelas} has the structure
\beqa
ds^2 \, & = & \, dr^2 \, +\, r^2 \, [
A\, (d\theta_1^{\, 2}\, +\, \sin^2 \theta_1\, d\phi_1^{\,2}\, +\, 
d\theta_2^{\, 2}\, +\, \sin^2 \theta_2\, d\phi_2^{\,2}) + \nonumber \\
&& + B\, (d\psi\, +\, \cos \theta_1 \, d\phi_1 \, +\, \cos \theta_2 \, 
d\phi_2)^{\, 2}]
\eeqa
In relating this to T-dual intersecting NS-branes, ref \cite{dm} suggest 
replacing the Maurer-Cartan forms of $\IS^2$ to those of $\IR^2$
$d\theta_1,\sin\theta_1 d\phi_1 \to dx_6,dx_7$ (and analogously for 
$\theta_2$, $\phi_2$) leading to
\beqa
ds^2 \, = \, dr^2\, + \, r^2\, [\,
A\, (dx_6^{\, 2} + dx_7^{\, 2} + dx_8^{\, 2} + dx_9^{\, 2}) \, + \,
B\, (d\psi \, +\, x_6 dx_7 \, +\, x_8 dx_9 )^{\,2}\,]
\label{conidm}
\eeqa
Intuitively, the non-supersymmetric conifold corresponds to a geometry 
with off diagonal metric in 6789 space, leading to rougly speaking
\beqa
ds^2 \, & = & \, dr^2 + \, r^2\, [\,
A\, (dx_6^{\, 2} + dx_7^{\, 2} + dx_8^{\, 2} + dx_9^{\, 2}
+ \alpha \, dx_6\, dx_8 \, +\, \beta\, dx_7\, dx_9) \, + \nonumber \\
&&+ B\, (d\psi \, +\, x_6 dx_7 \, +\, x_8 dx_9 )^{\, 2}\,]
\eeqa
For $SU(2)$ angles, which amounts to $\alpha=-\beta$, the metric is 
equivalent to (\ref{conidm}) via a coordinate change. For non-$SU(2)$ 
angles, susy is broken even in the asymptotic region. The metric must 
relax towards a susy metric with some asymptotic off-diagonal metric in 
6789. This is provided by a deformed conifold metric
\cite{candelas, mt}, which can be written (using $\IR^2$ Maurer-Cartan 
forms) as \cite{ohtayokono}
\beqa
ds^2 \, & = & \, dr^2 + \, [ \,
C(r)\, (dx_6^{\, 2} + dx_7^{\, 2} + dx_8^{\, 2} + dx_9^{\, 2}) \, + \,
D(r)\, (dx_6,dx_7)\, \cdot\,  R(\psi)\,  \cdot\, (dx_8,dx_9)^T \, + 
\nonumber 
\\
&&+ E(r)\, (d\psi \, +\, x_6 dx_7 \, +\, x_8 dx_9 )\, ]
\eeqa
where $R(\psi)$ is a $2\times 2$ rotation matrix with angle $\psi$. Hence 
the non-supersymmetric conifold likes to develop a deformed conifold at its 
tip, since it is the Calabi-Yau metric with more similar asymptotic 
behaviour \footnote{T-duality of deformed conifolds and recombined 
branes has appeared in \cite{diamond,ohtayokono, radu2}}. 
Even though the argument is suggestive, a detailed match 
of features like the $\psi$ dependence in the off-diagonal metric in 6789 
would require being able to treat localized sources in the T-dual NS-brane 
configuration.

\subsection{Discussion of compact models}

In several respects the above instabilities are similar to those arising
from closed string twisted tachyons at non-supersymmetric orbifold 
singularities \cite{aps}. Namely, our instabilities are localized at 
singular points, they signal the dynamical resolution of the singularity, 
and for large resolutions are triggered by gravitational interactions 
(while at short distances they have a stringy/ M-theory origin). Also, in 
non-compact setups the dynamical smoothing of singularities proceeds to 
infinity in both cases, with an expandind shell of energy separating the 
(possibly partially) smoothed region from the still non-supersymmetric 
asymptotic one.

In this section we would like to point out that nevertheless the behaviour 
of both instabilities seems qualitatively different in compact setups. It 
has been argued from different viewpoints \cite{aps,vafa} that 
Zamolodchikov's c-theorem implies that in compact setups condensation of 
closed string twisted tachyons must lead to a reduction of the number of 
spacetime dimensions. Such an argument is clearly not applicable to our 
instabilities, since they arise in M-theory or in string theory in the 
regime where the world-sheet CFT description breaks down (namely at 
Strominger's conifold point \cite{strominger}, and analogs). 

In fact, it is easy to cook up configurations of D6-branes in compact 
setups which lead to compact threefolds with instabilities of the kind 
discussed above, and which relax to a supersymmetric situations without 
loss of spacetime dimension. A simple and controllable situation is to 
start with D6-branes wrapped on factorizable special lagrangian 2-cycles 
on $\IT^2\times \IT^2$, and slightly change the $\IT^2$ complex structures. 
The D6-branes suffer a slight recombination after which they wrap a 
recombined 2-cycle, which is special lagrangian in the new complex 
structure. From the viewpoint of M-theory, the lift after the complex 
structure deformation is a threefold with a number of conifold singularities 
of the above non-supersymmetric kind. Their corresponding instabilities 
trigger a dynamical deformation of the conifolds which ends at finite 
distance in the deformation parameters \footnote{As discussed in appendix 
\ref{kahler}, one must require models with several intersections, in order 
to avoid topological obstructions to the transition.}.

Namely, the field controlling the 3-cycle size reaches a minimum of its 
potential. In this sense, the configurations we have studied are the 
simplest cases of localized instabilities with no open string origin 
which have non-trivial minima from the unstable potential at finite 
distance. We hope our results, although qualitative, are inspiring in the 
search for other situations with these features.

\section{One- and three-angle systems}
\label{onethree}

In this section we perform a similar analysis in other systems of 
D6-branes at angles. Even though we encounter some new features, the main 
ideas are analogous, so our discussion is more sketchy.

\subsection{One angle system}
\label{oneangle}

Consider a system of two D6-branes intersecting over a six-dimensional 
subspace of their volumes. The geometry is as follows
\beqa
{\rm D6}\;\; 0\; 1\; 2\; 3\; 4\; \times \; 6\; \times \; [\; 8\; 9\; 
]_{\theta} \nonumber \\
{\rm D6}'\;\; 0\; 1\; 2\; 3\; 4\; \times \; 6\; \times \; [\; 8\; 9\; 
]_{\theta'} 
\eeqa
Contrary to Section \ref{twoangle}, this system is non-supersymmetric for 
any non-zero value for $\Delta \theta=\theta-\theta'$. The open string 
sector at the intersection always contains a tachyon, with mass
\beqa
\alpha' M^2\, =\, -\frac 1{2\pi} \Delta\theta 
\eeqa
which triggers the recombination of the intersecting D6-branes. In a 
non-compact setup the recombination proceeds to infinity, while for 
D6-branes wrapped on 1-cycles $(n_8,n_9)$, $(n_8',n_9')$ on $\IT^2$, the 
recombined D6-brane ends up wrapping the cycle $(n_8+n_8',n_9+n_9')$, the 
minimal volume cycle in its homology class. The final state is 1/2 BPS.

\medskip

By separating the branes in the transverse directions 5, 7, the weakly 
interacting D6-branes can be reliable lifted to M-theory as a 
five-dimensional metric given by a superposition of two Taub-NUT spaces.
In this regime, in a compact setup, the 6d tension of the system is the 
addition of the individual tensions
\beqa
Z_{\rm total}\, =\, Z_{n_8,n_9}\, +\, Z_{n_8',n_9'}\, =\, T_{KK}\, 
[\, [\, (n_8 R_8)^2\, +\, (n_9 R_9)^2\,]^{1/2}\, +\, [\, (n_8' R_8)^2\, 
+\, (n_9' R_9)^2\,]^{1/2} \, ] \nonumber
\eeqa
This is always larger than the BPS bound for a state with those charges, 
which can be readily computed to be
\beqa
Z_{BPS} =  T_{KK} \, [\, (\, n_8\, +\, n_8'\,)^2\,  R_8^{\, 2}\, +\, 
(\, n_9\, +\, n_9'\,)^2 R_9^{\,2}\, ]^{1/2}
\eeqa
The BPS bound is saturated by a Taub-NUT wrapped on the 1-cycle 
$(n_8+n_8',n_9+n_9')$. Therefore, the system of two Taub-NUT's well 
separated in the directions 57 is unstable against decay to a lower 
energy configuration. 

We propose this decay to occur in a manner similar to Section 
\ref{twoangle}. Namely, far in the Coulomb branch gravitational 
interactions between the Taub-NUT cores leads to their approach until
reaching Planckian distances. At this stage an instability develops, 
localized near the core intersections. This instability triggers the 
recombination of their cores, ending up in a geometry which is a Taub-NUT 
$\IS^1$ fibration degenerating over the 1-cycle $(n_8+n_8',n_9+n_9')$.
 
Interestingly enough, the process represents the decay of a 
non-supersymmetric five-dimensional non-trivial geometry into a 
supersymmetric factorized geometry of the form $\IX_4\times \IS^1$, where 
$\IX_4$ is the final Taub-NUT geometry, and the $\IS^1$ spans the 
direction transverse to $(n_8+n_8',n_9+n_9')$ in $\IT^2$.

\begin{figure}
\begin{center}
\centering
\epsfysize=3cm
\leavevmode
\epsfbox{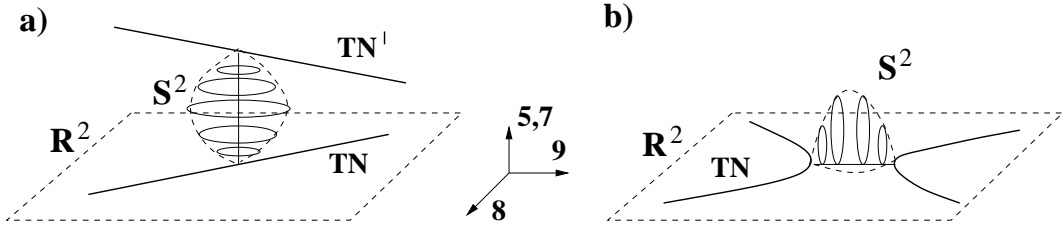}
\end{center}
\caption[]{\small 
Schematic picture of the lift of configurations of D6-branes intersecting 
at one angle on the Coulomb (a) and Higgs (b) branch. We have highlighted the 
two different non-trivial two-spheres in these geometries, which are 
obtained as $\IS^1$ fibrations over two different segments on the base.}
\label{flop}
\end{figure}

From the local viewpoint, the geometrical transition is a topology 
changing transition, in which a 2-sphere present in the Coulomb branch 
shrinks, and a new topologically different 2-sphere grows in the Higgs 
branch, see Fig~\ref{flop}. 

\subsection{Three angle case}
\label{threeangle}

Consider the case of two D6-branes intersecting over a four-dimensional 
subspace of their volume. The geometry is
\beqa
{\rm D6}\;\; 0\; 1\; 2\; 3\; [\; 4\; 5\; ]_{\theta_1}\; 
[\; 6\; 7\; ]_{\theta_2}\; [\; 8\; 9\; ]_{\theta_3} 
\nonumber \\
{\rm D6'}\;\; 0\; 1\; 2\; 3\; [\; 4\; 5\; ]_{\theta_1'} \;
[\; 6\; 7\; ]_{\theta_2'} \; [\; 8\; 9\; ]_{\theta_3'}
\eeqa

The configuration is supersymmetric when the $SO(6)$ rotation between the 
two spanned 3-planes lies within an $SU(3)$ subgroup, i.e.
\beqa
\Delta\theta_1 \pm \Delta\theta_2 \pm \Delta\theta_3 = 0
\quad \quad {\rm for}\; {\rm some}\; {\rm choice}\; {\rm of}\; 
{\rm signs}. 
\eeqa
The intersection contains a chiral multiplet (arising from D6-D6$'$ open 
strings) whose scalar component vev parametrizes recombination of 
3-cycles into a single smooth special lagrangian 3-cycle. In contrast with 
previous situations, there is no Coulomb branch for this system, due to 
the absence of overall transverse dimensions.

In non-supersymmetric situations this complex scalar may be tachyonic or 
non-tachyonic. Assuming $0\leq \Delta \theta_i \leq \pi$, the lightest 
scalars have masses
\beqa
& \alpha' M^2 \, =\,  \frac 1{2\pi} (-\Delta\theta_1 + \Delta\theta_2 + 
\Delta\theta_3) \quad \quad 
& \alpha' M^2 \, =\,  \frac 1{2\pi} (\Delta\theta_1 - \Delta\theta_2 + 
\Delta\theta_3) \nonumber \\
& \alpha' M^2 \, =\,  \frac 1{2\pi} (\Delta\theta_1 + \Delta\theta_2 -
\Delta\theta_3) \quad \quad 
& \alpha' M^2 \, =\,  -\frac 1{2\pi} (-\pi +\Delta\theta_1 - 
\Delta\theta_2 + \Delta\theta_3)  \nonumber 
\eeqa
In angle space, the non-tachyonic range lies within a tetrahedron 
introduced in \cite{imr}, see \cite{raultoron} for further details.

For angles in the tachyonic range, there exist manifolds with smaller
volume (in fact special lagrangian manifolds) and same asymptotic 
behaviour \cite{angle}. Hence, the tachyon triggers decay to a single 
D6-brane wrapped on such 3-manifold, which is a 1/8 BPS state. In 
non-compact setups the recombination proceeds to infinity, via the 
familiar shell expansion process, while in compact ones it stops at the 
minimal volume cycle in its homology class. 

For non-tachyonic angles, such smaller volume 3-cycles do not exist, 
hence the configuration remains non-supersymmetric, but stable against 
small perturbations. In compact setups, however, non-perturbative global 
rearrangements of D6-branes may allow decay of non-tachyonic 
configurations \footnote{In fact it is easy to describe a domain wall 
which interpolates between two such (meta)stable minima. It is 
given by a D8-brane spanning 012456789, and at $x^3=0$. Consider a 
configuration of (semi-infinite) D6-branes wrapped on 3-cycles 
$[\Pi_a]$ and spanning 012 and $x^3<0$, and ending on the D8-brane; and a 
similar configuration of D6-branes wrapped on cycles $[\Pi_a']$ and 
spanning 012 and $x_3>0$. Nucleation of these D8-${\ov D8}$ domain walls 
(or expansion of a spherical D8-brane) mediates the decay of metastable 
vacua, such as those in \cite{intersecting}. For supersymmetric cases 
\cite{orbif,susy} the D8-brane domain wall is BPS, and separates different 
$N=1$ susy configurations.}.

\medskip

Let us now describe the lift of these systems. As in Section \ref{instab}
it is convenient to discuss the supersymmetric case first. This lift must
correspond \cite{jaume} to a $G_2$-holonomy seven-dimensional singularity. 
In the near core regime, or infinite asymptotic radius limit, the metric 
(and its relation to intersecting D6-branes) has been explicitly 
discussed in \cite{aw} (see also \cite{klaus}), which moreover shows that 
the resolution of the singularity corresponds to the D6-brane recombination. 

Concerning the non-supersymmetric case, it is possible to perform a BPS 
analysis by considering objects wrapped on 3-cycles on $\IT^6$ \footnote{
Given the absence of overall transverse dimensions, cancellation of charge 
in the compact setup should be required for consistency; we will consider 
our two-brane system to be part of a larger set, and center on the 
dynamics of a particular instability, essentially unaffected by the 
rest.}. However, a first difficulty is that in the absence of Coulomb 
branch there  is not enough protection against strong corrections in 
lifting the initial intersecting D6-brane configuration. 

For the sake of the argument we could assume that the lift gives some 
kind of intersecting Taub-NUT metric, namely a non-$G_2$ metric in the 
topology of the singularity in \cite{aw}. Being non-supersymmetric, this 
geometry does not saturate the BPS bound. In the tachyonic range of 
angles, the BPS bound is however saturated by the state which gives the 
lift of the recombined D6-brane system (which is BPS, and hence must 
exist in M-theory). Hence the singularity is unstable against dynamical 
smoothing, which drives the configuration into the Higgs branch, where it 
eventually looks like a Taub-NUT fibration over the recombined 3-cycle.
In the non-tachyonic range of angles, there is no guarantee on the 
existence of BPS saturating objects, hence the stability of the 
singularity is unclear (namely, it may remain stable, or decay into some 
other different non-BPS geometry, stable at large $R_{10}$).

\section{Conclusions}
\label{conclusions}

In this note we have discussed the M-theory lift of the dynamics of 
diverse intersecting D6-brane systems. They provide interesting dynamical 
processes of purely gravitational systems in string / M-theory. Given the 
difficult regime where such processes take place, our main tool has been 
an analysis of energetics using BPS formulae. Several questions concerning 
the explicit description of the geometries involved are beyond our tools.

A possible improvement in this respect could be provided by studying the 
supergravity solutions for intersecting/recombined NS-brane systems, and 
using T-duality to uncover the intersecting/recombined Taub-NUT metrics in 
our discussions. Unfortunately, most metrics for intersecting brane 
systems in the literature involve smeared sources, whereas our purposes
would require localized source solutions. These have become available more 
recently \cite{smith} and we may expect some progress in this direction.

Clearly many generalizations of our results are possible, in particular 
involving singularities resulting from lifts of more D6-branes (for 
instance, threefold singularities $xy=z^n w^m$ from two bunches of $n$, 
$m$ D6-branes, etc), or involving D6-branes and O6-planes. We hope these 
and other examples are helpful in extending our picture of condensation of 
instabilities associated to non-supersymmetric singularities.

\centerline{\bf Acknowledgements}

I thank G.~Aldazabal, L.~E.~Ib\'a\~nez, R.~Rabad\'an, and E.~Silverstein 
for useful discussions. I also thank M.~Gonz\'alez for kind support and 
encouragement. This work is supported by the Ministerio de Ciencia y 
Tecnolog\'{\i}a (Spain) through a Ram\'on y Cajal contract.

\bigskip

\appendix

\section{Recombined special lagrangian cycles}
\label{slag}

Following \cite{angle}, we review the construction of special lagrangian
2-cycle mediating the recombination of two intersecting 2-planes. For the 
$n$-cycle case, see \cite{angle}. The 2-cycles considered are of the form
\beqa
(z_1(t,\varphi), z_2(t,\varphi)) \, = (\cos\varphi\, z_1(t),\sin\varphi \,
z_2(t))
\eeqa
with $z_i$ giving coordinates in $\IR^4=\IR^2\times \IR^2$, and $\phi,t$ 
parametrizing the 2-cycle. The fact that the 2-cycle is special lagrangian 
follows from
\beqa
\frac{dz_1}{dt}= i {\ov z}_2 \quad ; \quad \frac{dz_2}{dt}= i {\ov z}_1 
\eeqa
The initial boundary condition is taken $z_i(t=0)=c_i\in \IR$.

For any given $t=t_0$, the region $|t|\leq |t_0|$, $\varphi$ arbitrary, 
defines a special lagrangian 2-cycle, with boundary two circles 
$z_i(t=\pm t_0,\varphi)$ lying within two 2-planes at relative angles 
$\Delta \theta_i= 2{\rm Arg}(z_i(t_0))$. These angles are in general not 
in $SU(2)$ relation, and moreover for any set of angles in non-$SU(2)$ 
relation there exists $t_0$ such that $\Delta \theta_i=2 {\rm Arg}(z_i(t_0))$.

The region provides a recombined special lagrangian 2-cycle with a 
recombination size parametrized by the $c_i$, and whose boundary can be 
exactly glued to two 2-planes intersecting at non-susy angles. Hence, 
given two D6-branes at non-supersymmetric angles, the above 2-regions 
provide recombined 2-cycles mediating the tachyon condensation. The decay
proceeds via a family of 2-cycles, with larger and larger $c_i$, and glued 
to the 2-planes at larger and larger radii, in a shell expansion picture. 
The expansion of the shell is triggered by the difference in tension 
between the supersymmetric interior region and the non-susy exterior.

\section{K\"ahler constraints in conifold transitions from D6-branes}
\label{kahler}

There is an interesting constraint \cite{candelas, gmv} on the possible 
conifold transitions that can be undergone by a compact Calabi-Yau with 
conifold singularities. Given a set of conifold singularities, the 
deformation phase is only possible if there are homology relations among 
the 2-spheres which shrink in the singular limit. 

In particular, this implies that a compact Calabi-Yau cannot undergo a 
transition at a single conifold point. This is argued as follows: starting 
in the resolution phase, the small 2-sphere has a compact dual 4-cycle 
intersecting the 2-cycle. After the transition to the putative deformation 
phase, the 4-cycle would become a 4-chain with boundary the 3-sphere. 
Hence the 3-sphere would be homologically trivial, and would allow for no 
modulus to parametrize the branch. On the other hand, with for instance 
two conifold points with homologically related shrinking 2-spheres, the 
dual 4-cycle intersects both; in the deformation phase it becomes a 
4-chain defining a homological relation between the two 3-spheres, hence 
allowing for one complex parameter moduli space. In general the dimension 
of the moduli space of deformations is given by the number of independent 
such homology relations.

\begin{figure}
\begin{center}
\centering
\epsfysize=5cm
\leavevmode
\epsfbox{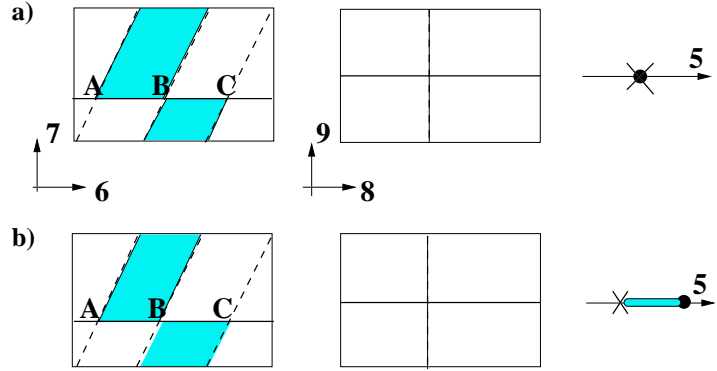}
\end{center}
\caption[]{\small
Intersecting D6-branes provide a skeleton picture for certain Calabi-Yau 
threefolds with conifold singularities. The shaded area provides the 
skeleton picture for a 3-chain defining a homology relation between 
2-spheres A, C, in the small resolution phase. This homology relaton 
allows a conifold transition involving the corresponding nodes.}
\label{homology}
\end{figure}       

This kind of constraint should also arise in Calabi-Yau manifolds 
obtained as lifts of configurations of intersecting D6-branes wrapped on 
(supersymmetric) 2-cycles. We would like to show that in fact they are 
easily obtained in terms of the latter. Indeed, the existence of the 
deformation phase involving a number of conifold singularities 
corresponds to entering the Higgs branch in which a number 
of D6-brane intersections are smoothed out by recombination. This is the 
Higgs branch of a gauge field theory with eight supercharges, living on 
the non-compact piece of the D6-brane world-volumes. Considering 
e.g. two D6-branes with $N$ intersections, it is a $U(1)$ gauge 
theory with $N$ charged hypermultiplets (this $U(1)$ is the difference of 
the D6-brane world-volume $U(1)$'s, and all hypers have the same charge).
The Higgs branch is absent for $N=1$, while for higher $N$ there is an
$N-1$ (quaternionic) dimensional Higgs branch, as follows from the F-term 
and D-term equations
\beqa
& Q_1 {\tilde Q}_1 + \ldots Q_N {\tilde Q}_N = 0 \nonumber \\
& |Q_1|^1 + \ldots |Q_N|^2 - |{\tilde Q}_1|^1 + \ldots |{\tilde Q}_N|^2 
= 0
\eeqa
where $(Q_i,{\tilde Q}_i)$ denotes the $i^{th}$ hypermultiplet. Particular 
mesonic directions in the Higgs branch are given by the vevs $Q_i={\tilde 
Q}_j=v$, $i\neq j$, with other fields set to zero
\footnote{In terms of D6-brane geometry, the existence of Higgs branch 
corresponds to the existence of a recombined Slag 2-cycle in the correct 
homology class. In a T-dual version (see e.g. \cite{ralph,raultoron}) it is 
related to the existence of certain gauge field connection carrying the 
appropriante Chern classes.}.

\begin{figure}
\begin{center}
\centering
\epsfysize=3cm
\leavevmode
\epsfbox{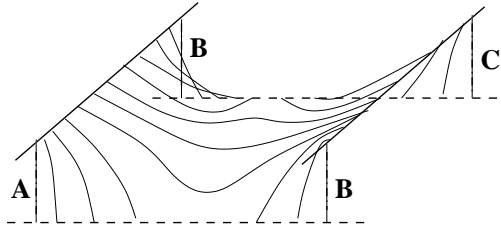}
\end{center}
\caption[]{\small
The 3-chain $\Sigma_3$ mentioned in figure \ref{homology} is obtained by 
fibering the M-theory circle over the region here depicted. Notice that 
the 2-spheres at the two locations B are glued onto each other, and do 
not belong to the boundary of $\Sigma_3$.}
\label{threecycle}
\end{figure}       

The D6-brane / gauge field theory analysis then suggests that in the 
M-theory lift the corresponding threefolds have homology relations among 
the 2-cycles at the conifold singularities associated to hypermultiplets 
getting a vev. In fact, the corresponding 3-chains are easily constructed 
using the D6-brane picture. In figure \ref{homology}a we have depicted an 
example with three intersections, in the singular conifold limit, while 
in figure \ref{homology}b separation of D6-branes in $x^5$ leads to a 
small resolution of the conifolds. The 2-spheres arise from fibering the 
M-theory circle over segments in $x^5$ and at locations A, B, C in 67 (and 
at the intersection in 89). A 3-chain $\Sigma_3$ defining a homology 
relation between the 2-spheres at A, C, is obtained by fibering the 
M-theory circle over a two-dimensional region whose projections are given 
by the shaded area. An improved picture of this region is provided in 
figure \ref{threecycle} (notice that the 2-sphere B is not a boundary of 
$\Sigma_3$). This homology relation allows the conifold transition 
associated to the mesonic brane involving the hypers at intersections A, 
C.


\begin{thebibliography}{99}

\bibitem{sen}
A.~Sen, `NonBPS states and Branes in string theory', hep-th/9904207.

\bibitem{aps}
A.~Adams, J.~Polchinski, E.~Silverstein, `Don't panic! Closed string 
tachyons in ALE space-times', JHEP 0110 (2001) 029, hep-th/0108075.

\bibitem{dabholkar}
A.~Dabholkar, `On condensation of closed string tachyons', hep-th/0109019.

\bibitem{vafa}
C.~Vafa, `Mirror symmetry and closed string tachyon condensation', 
hep-th/0111051.

\bibitem{others}
A.~Dabholkar, `Tachyon condensation and black hole entropy',
Phys. Rev. Lett. 88 (2002) 091301, hep-th/0111004; \\
J.~A.~Harvey, D.~Kutasov, E.~J.~Martinec, G.~Moore, `Localized tachyons 
and RG flows', hep-th/0111154; \\
A.~Dabholkar, C.~Vafa, `tt* geometry and closed string tachyon potential',
JHEP 0202 (2002) 008, hep-th/0111155; \\
J.~R. David, M.~Gutperle, M.~Headrick, S.~Minwalla, `Closed string tachyon 
condensation on twisted circles', JHEP 0202 (2002) 041, hep-th/0111212; \\
S.-K.~Nam, S.-J.~Sin, `Condensation of localized tachyons and space-time 
supersymmetry', hep-th/0201132; \\
A.~Font, A.~Hern\'andez, `Non-supersymmetric orbifolds', hep-th/0202057; 
\\
S.~-J.~Sin, `Tachyon mass, c function and counting localized 
degrees of freedom', hep-th/0202097.

\bibitem{fluxbranes}
F.~Dowker, J.~P.~Gauntlett, G.~W.~Gibbons, G.~T.~Horowitz, `The Decay of 
magnetic fields in Kaluza-Klein theory', Phys. Rev. D52 (1995) 6929, 
hep-th/9507143; `Nucleation of p-branes and fundamental strings',
Phys. Rev. D53 (1996) 7115, hep-th/9512154; \\
M.~S.~Costa, M.~Gutperle, `The Kaluza-Klein Melvin solution in M theory',
JHEP 0103 (2001) 027, hep-th/0012072; \\
J.~G.~Russo, A.~A.~Tseytlin, 
`Magnetic flux tube models in superstring theory'
Nucl. Phys. B461 (1996) 131, hep-th/9508068;
`Green-Schwarz superstring action in a curved magnetic Ramond-Ramond 
background', JHEP 9804 (1998) 014, hep-th/9804076;
`Magnetic backgrounds and tachyonic instabilities in closed superstring 
theory and M theory', Nucl. Phys. B611 (2001) 93, hep-th/0104238; 
`Supersymmetric fluxbrane intersections and closed string tachyons',
JHEP 0111 (2001) 065, hep-th/0110107; \\
T.~Suyama, `Closed string tachyons in non-supersymmetric heterotic 
theories', JHEP 0108 (2001) 037, hep-th/0106079;
`Melvin background in heterotic theories',
Nucl. Phys. B621 (2002) 235, hep-th/0107116;
`Properties of string theory on Kaluza-Klein Melvin background',
hep-th/0110077; 
`Charged tachyons and gauge symmetry breaking', JHEP 0202 (2002) 033, 
hep-th/0112101; \\
T.~Takayanagi, T.~Uesugi, `Orbifolds as Melvin geometry', hep-th/0110099; 
\\
Y.~Michishita, P.~Yi, `D-brane probe and closed string tachyons',
hep-th/0111199.
 
\bibitem{fullclosed}
O.~Bergman, M.~R.~Gaberdiel, `Dualities of type 0 strings',
JHEP 9907 (1999) 022, hep-th/9906055; \\
S.~P.~De Alwis, A.~T.~Flournoy, `Closed string tachyons and semiclassical 
instabilities', hep-th/0201185.

\bibitem{rauljoan}
R.~Rabad\'an, J.~Sim\'on, `M-theory lift of brane-antibrane systems and 
localised closed string tachyons', hep-th/0203243.

\bibitem{orbif}
R.~Blumenhagen, L.~Gorlich, B.~Kors, `Supersymmetric 4-D orientifolds of 
type IIA with D6-branes at angles', JHEP 0001 (2000) 040, hep-th/9912204; \\
S.~Forste, G.~Honecker, R.~Schreyer, `Supersymmetric Z(N) x Z(M) 
orientifolds in 4-D with D branes at angles', Nucl. Phys. B593 (2001) 
127, hep-th/0008250.

\bibitem{intersecting}
R.~Blumenhagen, L.~Goerlich, B.~Kors, D.~Lust, `Noncommutative 
compactifications of type I strings on tori with magnetic background 
flux', JHEP 0010 (2000) 006, hep-th/0007024; \\
G.~Aldazabal, S.~Franco, L.~E.~Ibanez, R.~Rabadan, A.~M.~Uranga, `D=4 
chiral string compactifications from intersecting branes',
J. Math. Phys. 42 (2001) 3103, hep-th/0011073; `Intersecting brane 
worlds', JHEP 0102 (2001) 047, hep-ph/0011132; \\
R.~Blumenhagen, B.~Kors, D.~Lust, `Type I strings with F flux and B flux',
JHEP 0102 (2001) 030, hep-th/0012156; \\
L.~E.~Ibanez, F.~Marchesano, R.~Rabadan, `Getting just the standard model 
at intersecting branes', JHEP 0111 (2001) 002, hep-th/0105155; \\
S.~Forste, G.~Honecker, R.~Schreyer, `Orientifolds with branes at angles',
JHEP 0106 (2001) 004, hep-th/0105208; \\
R.~Blumenhagen, B.~Kors, D.~Lust, T.~Ott, `The standard model from stable 
intersecting brane world orbifolds', Nucl. Phys. B616 (2001) 3, 
hep-th/0107138; \\
D.~Cremades, L.~E.~Ibanez, F.~Marchesano, `SUSY Quivers, Intersecting 
Branes and the Modest Hierarchy Problem', hep-th/0201205, `Intersecting 
brane models of particle physics and the Higgs mechanism', hep-th/0203160.

\bibitem{susy}
M.~Cvetic, G.~Shiu, A.~M.~Uranga, `Chiral four-dimensional N=1  
supersymmetric type 2A orientifolds from intersecting D6 branes',
Nucl. Phys. B615 (2001) 3, hep-th/0107166; `Three family supersymmetric 
standard - like models from intersecting brane worlds', Phys. Rev. Lett. 
87 (2001) 201801, hep-th/0107143; `Chiral type II orientifold 
constructions as M theory on G(2) holonomy spaces', hep-th/0111179.

\bibitem{angle}
G.~Lawlor, `The angle criterion', Invent. Math. 95 (1989) 437.

\bibitem{bdl}
M.~Berkooz, M.~R.~Douglas, R.~G.~Leigh, `Branes intersecting at angles', 
Nucl. Phys. B480 (1996) 265, hep-th/9606139.

\bibitem{raulcosmo}
J.~Garcia-Bellido, R.~Rabadan, F.~Zamora, `Inflationary scenarios from 
branes at angles', JHEP 0201 (2002) 036, hep-th/0112147.

\bibitem{herdeiro}
C.~Herdeiro, S.~Hirano, R.~Kallosh, `String theory and hybrid inflation / 
acceleration', JHEP 0112 (2001) 027, hep-th/0110271.

\bibitem{op}
N.~A.~Obers, B.~Pioline, `U duality and M theory', Phys. Rept. 318 (1999) 
113, hep-th/9809039.

\bibitem{radu1}
R.~de Mello Koch, K.~Oh, R.~Tatar, `Moduli Space for Conifolds as 
Intersection of Orthogonal D6 branes', Nucl.Phys. B555 (1999) 457,
hep-th/9812097.

\bibitem{uraconi}
A.~M.~Uranga, `Brane configurations for branes at conifolds', JHEP 9901 
(1999) 022, hep-th/9811004.

\bibitem{dm}
K.~Dasgupta, S.~Mukhi, `Brane constructions, conifolds and M theory',
Nucl. Phys. B551 (1999) 204, hep-th/9811139.

\bibitem{candelas}
P.~Candelas, P.~S.~Green, T.~Hubsch, Phys. Rev. Lett. 62 (1989) 1956;
Nucl. Phys. B330 (1990) 49; \\
P.~Candelas, X.~C.~de la Ossa, `Comments on conifolds',
Nucl. Phys. B342 (1990) 246.

\bibitem{gms}
B.~R.~Greene, D.~R.~Morrison, A.~Strominger, `Black hole condensation and 
the unification of string vacua', Nucl. Phys. B451 (1995) 109, 
hep-th/9504145.

\bibitem{tomas}
E.~Bergshoeff, B.~Janssen, T.~Ortin, `Kaluza-Klein monopoles and gauged 
sigma models', Phys. Lett. B410 (1997) 131, hep-th/9706117.

\bibitem{imamura}
Y.~Imamura, `Born-Infeld action and Chern-Simons term from Kaluza-Klein 
monopole in M theory', Phys. Lett. B414 (1997) 242, hep-th/9706144.

\bibitem{mt}
R.~Minasian, D.~Tsimpis, `On the geometry of nontrivially embedded 
branes', Nucl. Phys. B572 (2000) 499, hep-th/9911042.

\bibitem{diamond}
M.~Aganagic, A.~Karch, D.~Lust, A.~Miemiec, `Mirror symmetries for brane 
configurations and branes at singularities', Nucl. Phys. 
B569 (2000) 277, hep-th/9903093.

\bibitem{ohtayokono}
K.~Ohta, T.~Yokono, `Deformation of conifold and intersecting branes',
JHEP 0002 (2000) 023, hep-th/9912266.

\bibitem{radu2}
K.~Dasgupta, K.~Oh, J.~Park, R.~Tatar, `Geometric Transition versus 
Cascading Solution', JHEP 0201 (2002) 031, hep-th/0110050.

\bibitem{strominger}
A.~Strominger, `Massless black holes and conifolds in string theory',
Nucl. Phys. B451 (1995) 96, hep-th/9504090.

\bibitem{imr}
L.~E.~Ibanez, F.~Marchesano, R.~Rabadan, as in \cite{intersecting}.

\bibitem{raultoron}
R.~Rabad\'an, `Branes at angles, torons, stability and supersymmetry',
Nucl. Phys. B620 (2002) 152, hep-th/0107036.

\bibitem{jaume}
J.~Gomis, `D-branes, holonomy and M theory', Nucl. Phys. B606 (2001) 3, 
hep-th/0103115.

\bibitem{aw}
M.~Atiyah, E.~Witten, `M theory dynamics on a manifold of G(2) holonomy', 
hep-th/0107177.

\bibitem{klaus}
K.~Behrndt, `Singular 7-manifolds with $G_2$ holonomy and intersecting 
6-branes', hep-th/0204061.

\bibitem{smith}
A.~Fayyazuddin, D.~J.~Smith, `Localized intersections of M5-branes and 
four-dimensional superconformal field theories', JHEP 9904 (1999) 030,
hep-th/9902210; `Warped AdS near-horizon geometry of completely localized  
intersections of M5-branes', JHEP 0010 (2000) 023, hep-th/0006060;
B.~Brinne, A.~Fayyazuddin, S.~Mukhopadhyay, D.~J.~Smith, `Supergravity 
M5-branes wrapped on Riemann surfaces and their QFT duals', JHEP 0012 
(2000) 013, hep-th/0009047; B.~Brinne, A.~Fayyazuddin, T.~Zehra Husain, 
D.~J.~Smith, `N=1 M5-brane geometries', JHEP 0103 (2001) 052, 
hep-th/0012194.

\bibitem{gmv}
B.~R.~Greene, D.~R.~Morrison, C.~Vafa, `A Geometric realization of 
confinement', Nucl. Phys. B481 (1996) 513, hep-th/9608039.

\bibitem{ralph}
R.~Blumenhagen, V.~Braun, R.~Helling, `Bound states of D(2p) D0 systems 
and supersymmetric p cycles', Phys. Lett. B510 (2001) 311, hep-th/0012157.


\end{thebibliography}
\end{document}